\documentclass[aip,sd,amsmath,amssymb,reprint]{revtex4-1}
\usepackage{graphicx}
\usepackage{multirow}
\usepackage{bm}

\begin{document}

\preprint{AIP/123-QED}

\title{Comment on ``Theoretical design of molecular nanomagnets for magnetic refrigeration'' [Appl. Phys. Lett. 103, 202410 (2013)]}

\author{Marco Evangelisti}
 \homepage{http://molchip.unizar.es/}
\affiliation{$^1$Instituto de Ciencia de Materiales de Arag\'{o}n (ICMA) and Departamento de F\'{\i}sica de la Materia Condensada,
CSIC -- University of Zaragoza, Pedro Cerbuna 12, 50009 Zaragoza, Spain}

\author{Giulia Lorusso}
 \homepage{http://molchip.unizar.es/}
\affiliation{$^1$Instituto de Ciencia de Materiales de Arag\'{o}n (ICMA) and Departamento de F\'{\i}sica de la Materia Condensada,
CSIC -- University of Zaragoza, Pedro Cerbuna 12, 50009 Zaragoza, Spain}%

\author{Elias Palacios}
\affiliation{$^1$Instituto de Ciencia de Materiales de Arag\'{o}n (ICMA) and Departamento de F\'{\i}sica de la Materia Condensada,
CSIC -- University of Zaragoza, Pedro Cerbuna 12, 50009 Zaragoza, Spain}%

\date{\today}

%

\maketitle

Garlatti {\it et al.}~\cite{Garlatti} report theoretical simulations aimed at showing that the best molecular nanomagnets (MNMs) for magnetic refrigeration between $T\simeq 10$~K and sub-Kelvin region are those made of strongly ferromagnetically-coupled magnetic ions. The authors make impeccable calculations leading to results that, apparently, contrast with the established belief in this research field.~\cite{Marco} We point out that the performance of any magnetic refrigerant is largely dependent on extrinsic parameters, viz., the experimental conditions that encompass the type of thermodynamic cycle employed for the refrigeration. The main conclusion of the title work is based on assuming that the refrigeration proceeds from 10~K down to 1~mK via a single stage, governed by the Carnot cycle. This Comment revises the results in Ref.~[\onlinecite{Garlatti}], while showing that the experimental conditions considered are impracticable and inconvenient.

By adiabatically demagnetizing a system of ideally non-interacting MNMs, the base temperature to be reached is $T_{\rm base}=T_{\rm hot}\cdot B_{\rm cold}/B_{\rm hot}$, where $T_{\rm hot}$ and $B_{\rm hot}$ are the initial temperature and applied field, respectively, and $B_{\rm cold}$ denotes the applied field at the end of the cooling procedure.~\cite{Pobell} By letting $B_{\rm cold}\rightarrow 0$, the system becomes sensitive to any perturbation and the previous expression should be replaced by $T_{\rm base}=T_{\rm hot}\cdot \sqrt{B_{\rm cold}^2+b^2}/B_{\rm hot}$, where $b$ is the internal field, which is determined by the magnetic anisotropy and magnetic interactions, e.g., dipole-dipole coupling. Intramolecular ferromagnetism maximizes the molecular spin $S$. The larger $S$, the likely stronger is the dipolar field that ultimately drives to a long-range magnetically-ordered state below a critical temperature $T_{\rm C}$, which typically occurs between 0.2~K and 0.8~K.~\cite{Fernando} Below $T_{\rm C}$, the magnetic entropy falls abruptly and so does the magnetocaloric effect (MCE), making MNMs not suited in principle for such low temperatures.~\cite{Marco} The vast majority of MNMs proposed as magnetic refrigerants are excellent candidates {\it limitedly} to temperatures between c.a. 1~K and 10~K, for which their MCE can be larger than that of conventional magnetic refrigerants, indeed. As noted by Garlatti {\it et al.},~\cite{Garlatti} spin dilution can be efficiently employed for suppressing intermolecular magnetic interactions, therefore permitting to attain lower temperatures. Unfortunately, the inherent downside is a dramatic reduction of the magnetic density, rendering this application useless for higher temperatures.~\cite{Marco}
\begin{figure}[b]
\includegraphics[width = 8 cm]{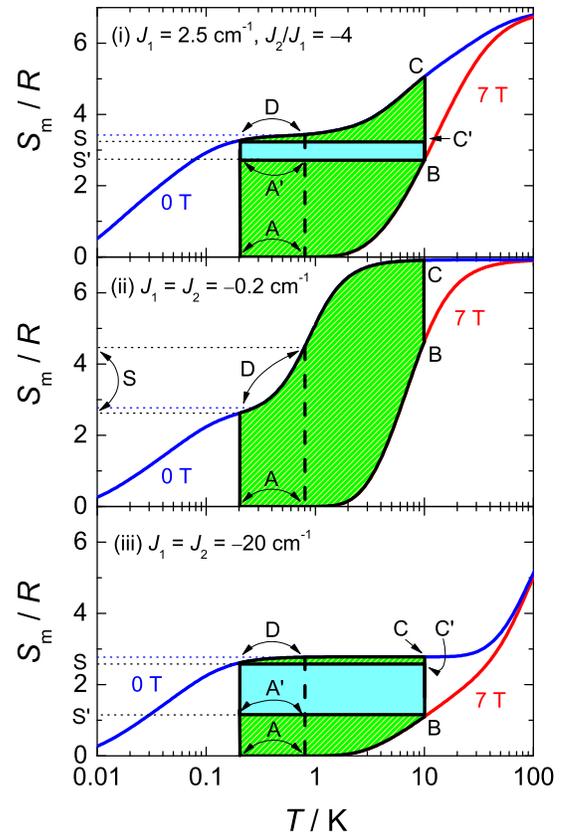}
\caption{\label{fig:1}
Magnetic entropies $S_0$ and $S_{\rm 7T}$, normalized to the gas constant $R$, for the three cases reported in Ref.~[\onlinecite{Garlatti}]. Idealized $S_0(T)$ curves for $b=0$ are reported. Carnot (A$^\prime$BC$^\prime$D) and Ericsson (ABCD) cycles, for $T_{\rm hot}=10$~K and $T_{\rm cold}$ either 0.3~K or 0.8~K, are depicted as shadow areas.}
\end{figure}

\begin{figure}[t!]
\includegraphics[width = 9 cm]{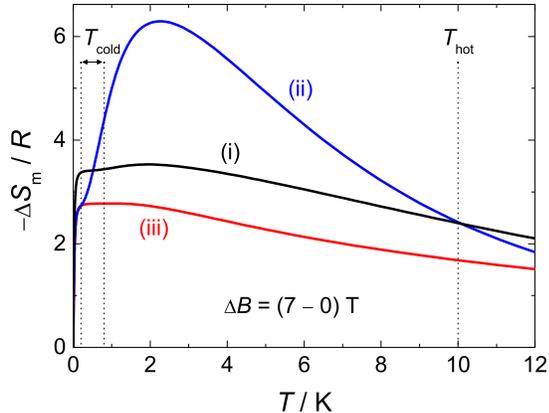}
\caption{\label{fig:2}
Magnetic entropy changes, normalized to the gas constant, for the three cases reported in Ref.~[\onlinecite{Garlatti}]. The applied field change is $\Delta B=B_{\rm hot}-B_{\rm cold}=(7-0)$~T, while the operating temperatures $T_{\rm hot}=10$~K and $T_{\rm cold}$ either 0.3~K or 0.8~K are indicated.}
\end{figure}

As in Ref.~[\onlinecite{Garlatti}], we consider the square-based pyramid with five $s=3/2$ spins interconnected by $J_1$ and $J_2$ exchanges, and we analyze the following three cases: (i) $J_1=2.5$~cm$^{-1}$, $J_2/J_1=-4$; (ii) $J_1=J_2=-0.2$~cm$^{-1}$; (iii) $J_1=J_2=-20$~cm$^{-1}$. Figure~\ref{fig:1} reproduces the same calculations of the magnetic entropy $S_{\rm m}$ (hereafter also denoted as $S_0$ for $B_{\rm cold}=0$ and $S_{\rm 7T}$ for $B_{\rm hot}=7$~T) that were in Figure~4 of Ref.~[\onlinecite{Garlatti}], though we replace the idealized $b=0$ by a nonzero value for the aforementioned reason. For the sake of simplicity, we assume the same $b=0.02$~T for the three cases. This value is comparable to that encountered in dipolar MNMs and, though small, it becomes relevant when $B_{\rm cold}=0$.~\cite{Fernando} Mimicking a real system and contrary to Ref.~[\onlinecite{Garlatti}], $b$ causes $S_0(T)$ to fall to zero for $T\rightarrow 0$ and inhibits $T$ to span down to $T_{\rm cold}=1$~mK also for (i) and (iii), assuming the same $T_{\rm hot}=10$~K and $\Delta B=B_{\rm hot}-B_{\rm cold}=(7-0)$~T that were in Ref.~[\onlinecite{Garlatti}]. Hereafter, we assign to $T_{\rm cold}$ either the value of 0.2~K or 0.8~K.~\cite{Fernando} Next, using the data in Fig.~\ref{fig:1}, we straightforwardly obtain the magnetic entropy changes $\Delta S_{\rm m}(T)=S_{\rm 7T}(T)-S_0(T)$ that we depict in Figure~\ref{fig:2}. There is a significantly larger $-\Delta S_{\rm m}(T)$ for $0.5~{\rm K}\lesssim T\lesssim 10~{\rm K}$ in (ii), i.e., the one characterized by the weakest $J_1$ and $J_2$. Since the spin centers in (ii) are almost decoupled already at such low temperatures, $-\Delta S_{\rm m}(T\simeq 2.2~{\rm K})$ approaches closely the full entropy content, i.e., $5R\ln(2s+1)\simeq 6.9 R$. Case (i), and especially (iii), show larger $-\Delta S_{\rm m}(T)$ than (ii) for high temperatures (not shown in Fig.~\ref{fig:2}), that is, where the exchange energies are of the same order as $k_{\rm B}T$. Competing interactions in (i) promote the relatively larger number of low-lying spin states that result in the larger values at the lowest $T$.

Since $T_{\rm cold}$ is expected between c.a. 0.2~K and 0.8~K, the Carnot cycles in Ref.~[\onlinecite{Garlatti}] become the A$^\prime$BC$^\prime$D cycles that we depict in Figure~\ref{fig:1} for (i) and (iii). As in Ref.~[\onlinecite{Garlatti}], no Carnot cycle can be implemented for (ii) under the experimental conditions considered. The heat absorbed by the refrigerant material during a single Carnot cycle is $Q_{\rm c}=T_{\rm cold}[S_0(T_{\rm cold})-S_{\rm 7T}(T_{\rm hot})]$, see Fig.~\ref{fig:1}. The larger $Q_{\rm c}$, the more is the advantage for the targeted application. Table~\ref{tab:1} reports the $Q_{\rm c}$ values corresponding to the three cases and $T_{\rm cold}$ considered. As in Ref.~[\onlinecite{Garlatti}], the Carnot cycles implemented for the ferromagnetic (iii) provide the largest values of $Q_{\rm c}$ for both $T_{\rm cold}$ temperatures. Table~\ref{tab:1} also shows the amount of work consumed during each Carnot cycle, i.e., $W=(T_{\rm hot}-T_{\rm cold})[S_0(T_{\rm cold})-S_{\rm 7T}(T_{\rm hot})]$.

\begin{table}[t]
\setlength{\tabcolsep}{9pt}
\renewcommand{\arraystretch}{1.2}
\begin{tabular}{cc|c|c||c|c|}
\cline{3-6}
& & \multicolumn{2}{ c|| }{$T_{\rm cold}=0.2$~K} & \multicolumn{2}{ c| }{$T_{\rm cold}=0.8$~K} \\ \cline{3-6}
& & $Q_{\rm c}$ & $W$ & $Q_{\rm c}$ & $W$ \\ \cline{1-6}
\multicolumn{1}{ |c| }{\multirow{2}{*}{(i)} } &
\multicolumn{1}{ |c| }{Carnot} & 0.8 & 39.9 & 3.3 & 37.4 \\ \cline{2-6}
\multicolumn{1}{ |c  }{}                        &
\multicolumn{1}{ |c| }{Ericsson} & 5.6 & 250.4 & 22.8 & 233.6 \\ \cline{1-6}
\multicolumn{1}{ |c| } {(ii)} & Ericsson & 4.5 & 357.4 & 29.9 & 339.3 \\ \cline{1-6}
\multicolumn{1}{ |c| }{\multirow{2}{*}{(iii)} } &
\multicolumn{1}{ |c| }{Carnot} & 2.3 & 114.9 & 9.4 & 107.8 \\ \cline{2-6}
\multicolumn{1}{ |c  }{}                        &
\multicolumn{1}{ |c| }{Ericsson} & 4.5 & 185.0 & 18.4 & 174.6 \\ \cline{1-6}
\end{tabular}
\caption{\label{tab:1}
For $\Delta B=(7-0)$~T, heat ($Q_{\rm c}$) absorbed from the MNM at $T_{\rm cold}$ and work ($W$) consumed to accomplish the thermodynamic cycle, for both $T_{\rm cold}$ considered, assuming the Carnot and Ericsson cycles depicted in Figure~\ref{fig:1}. For the Carnot cycle, $Q_{\rm c}$ and $W$ are the areas S$^\prime$A$^\prime$DS and A$^\prime$BC$^\prime$D, respectively. For the Ericsson cycle, $Q_{\rm c}$ and $W$ are the areas 0ADS and ABCD, respectively. Both $Q_{\rm c}$ and $W$ are expressed in J~mol$^{-1}$.}
\end{table}

We criticize the choice of the thermodynamic refrigeration cycle adopted in Ref.~[\onlinecite{Garlatti}], at least for (i) and (ii), since the functionality of these hypothetical materials is far from being fully exploited with the aforementioned Carnot cycles. In a Ericsson cycle (ABCD in Fig.~\ref{fig:1}), on the contrary, the field changes isothermally, therefore taking full advantage of the shape of the $S_{\rm m}(T)$ curves, between $T_{\rm cold}$ and $T_{\rm hot}$. Table~\ref{tab:1} shows the values of $Q_{\rm c}$ and $W$ for the Ericsson cycles, where $Q_{\rm c}=-T_{\rm cold}\Delta S_{\rm m}(T_{\rm cold})$ and $W=-\int_{T_{\rm cold}}^{T_{\rm hot}}\Delta S_{\rm m}(T){\rm d}T$. What can be seen is that the Ericsson cycles provide significantly larger $Q_{\rm c}$ values. Looking at Fig.~\ref{fig:2}, we conclude that (i) and (ii) can refrigerate more than (iii) below 10~K and, specifically, (ii) is ideally suited for a $T_{\rm cold}$ between c.a. 1.5~K and 4~K because of the prominent $-\Delta S_{\rm m}$ maximum. Note that the performance of the refrigeration cycles, i.e. $Q_{\rm c}/W$, is c.a. 2~\% for $T_{\rm cold}=0.2$~K and 9-10~\% for $T_{\rm cold}=0.8$~K, irrespectively of the choice of the material and the type of refrigeration cycle.

Working with Ericsson cycles implies using thermal regeneration, which is the easier to implement the narrower is the temperature span of the refrigeration cycle. Besides, small thermal gradients are desirable in order to minimize irreversible heat flows and are beneficial for temperature stabilization at low temperatures. A common strategy to engineer an adiabatic demagnetization refrigerator for very low temperatures is by combining multiple cooling stages.~\cite{Pobell} Liquid $^4$He or a 4K-cryocooler is employed for (pre)cooling down to c.a. 4.2~K. From there, $T$ is lowered down to $1-1.5$~K either by pumping on $^4$He or by exploiting the functionality of a magnetic refrigerant. Within the sub-Kelvin region, refrigerating magnetically with diluted spins, such as in paramagnetic salts, permits attaining mK temperatures. Starting from such low $T$, magnetic refrigeration using nuclear magnetic moments can be applied for getting even closer to absolute zero. The gist is that, below liquid-$^4$He temperature, every cooling stage operates within a relatively narrow temperature drop, allowing the refrigerator to remain cold for a long time. Therefore, compatibly with the target $T_{\rm cold}$ and $\Delta B$, $-\Delta S_{\rm m}(T_{\rm cold})$ should be maximized by, e.g., playing with the magnetic interactions.

In conclusion, we welcome the results reported in Ref.~[\onlinecite{Garlatti}], although we disagree on their interpretation. We emphasize that, in order to rank a magnetic refrigerant as the {\it best} one, we should first define common experimental conditions among all contenders. A hypothetical realization of an adiabatic demagnetization refrigerator operating at liquid-$^4$He temperatures includes: case (ii), which is the only example presented, worth of consideration for an application near 4.2~K, eventually combined with a dilution of (i) or (iii) for temperatures lower thank 1~K. We believe that the field of sub-Kelvin magnetic refrigeration with MNMs is still largely unexplored, both theoretically as well as experimentally. For a successful recipe at such low temperatures, the intermolecular interaction and magnetic anisotropy should also be taken into account.
\\
\\
\indent We acknowledge financial support by MINECO through grant MAT2012-38318-C03-01.

\end{document}